\documentstyle[ltwol]{article}

\def\Journal#1#2#3#4{{#1} {\bf #2}, #3 (#4)}

\def\NPB{{\em Nucl. Phys.} B}
\def\PLB{{\em Phys. Lett.}  B}
\def\PRL{\em Phys. Rev. Lett.}
\def\PRD{{\em Phys. Rev.} D}

\begin{document}
\title{NON--PERTURBATIVE RESULTS IN GLOBAL SUSY AND TOPOLOGICAL FIELD 
THEORIES}

\author{D. BELLISAI}

\address{Universit\`a di Roma ``Tor Vergata", Via della Ricerca Scientifica,  
I-00133 Roma, Italy \\E-mail: bellisai@roma2.infn.it}

\author{F. FUCITO, A. TANZINI,  G. TRAVAGLINI}

\address{I.N.F.N.- Sezione di Roma II ``Tor Vergata", 
Via della Ricerca Scientifica,  I-00133 Roma, Italy \\E-mail: 
fucito@roma2.infn.it, tanzini@roma2.infn.it, 
travaglini@roma2.infn.it}


\twocolumn[\maketitle\abstracts{In this lecture we 
briefly review the Seiberg--Witten model  and 
explore some of its topological aspects.}]


\def\beqa{\begin{eqnarray}}
\def\eeqa{\end{eqnarray}}
\def\beq{\begin{equation}}
\def\eeq{\end{equation}}
\def\beqal{\begin{eqnarray}\label}
\def\beql{\begin{equation}\label}

\def\vol{\int d^4x\,\sqrt{-g}} 
\def\tvol{\int d^4x\,\sqrt{-\tilde{g}}} 
\def\grav{-\frac{1}{16 \pi }}
\def\half{\frac{1}{2}}
\def\gu{g^{\m\n}}
\def\gd{g_{\m\n}}
\def\tgd{\tilde{g}_{\m\n}}
\def\tgu{\tilde{g}^{\m\n}}
\def\lc{\raisebox{-.7ex}{$\stackrel{\textstyle <}{\sim}$}}
\def\gc{\raisebox{-.7ex}{$\stackrel{\textstyle >}{\sim}$}}
\def\R{\mbox{\rm I\kern-.18em R}}
\def\P{\mbox{\rm I\kern-.18em P}}
\def\Ds{\ {\big / \kern-.70em D}}
\def\uno{{1 \kern-.30em 1}}
\def\ds{\big / \kern-.90em {\ \p} }

\def\cDs{\ {\big / \kern-.70em {\cal D}}}
\def\Z{{Z \kern-.45em Z}}
\def\Q{{\kern .1em {\raise .47ex \hbox{$\scriptscriptstyle |$}}
\kern -.35em {\rm Q}}}
\def\de{\mbox{\it d}}
\def\Tr{\mbox{\rm Tr}}
\def\Im{\mbox{\rm Im}}
\def\cl{\mbox{\scriptstyle cl}}
\def\tr{\mbox{\rm tr}}
\def\ap{\alpha^{\prime}}
\def\psb{\bar{\psi}}
\def\chb{\bar{\chi}}
\def\lb{\bar{\lambda}}
\def\epsb{\bar{\varepsilon}}
\def\sb{\bar{\sigma}}
\def\ad{\dot{\alpha}}
\def\bd{\dot{\beta}}
\def\wb{\bar{w}}
\def\vb{\bar{v}}
\def\ib{\bar\imath}
\def\jb{\bar\jmath}
\def\um{^{\m}}
\def\un{^{\n}}
\def\dm{_{\m}}  
\def\dn{_{\n}}
\def\umn{^{\m\n}}
\def\dmn{_{\m\n}}
\def\umnrs{^{\m\n\r\s}}
\def\dmnrs{_{\m\n\r\s}}
\def\ua{^{\al}}  
\def\ub{^{\b}}
\def\da{_{\al}}
\def\db{_{\b}}
\def\ug{^{\g}}
\def\dg{_{\g}}
\def\uam{^{\al\m}}
\def\uan{^{\al\n}}
\def\uab{^{\al\b}}
\def\dab{_{\al\b}}
\def\dabgd{_{\al\b\g\d}}
\def\uabgd{^{\al\b\g\d}}
\def\udeab{^{;\al\b}}
\def\ddeab{_{;\al\b}}
\def\ddemunu{_{;\m\n}}
\def\udemunu{^{;\m\n}}
\def\ddemu{_{;\m}}  \def\udemu{^{;\m}}
\def\ddenu{_{;\n}}  \def\udenu{^{;\n}}
\def\ddea{_{;\al}}  \def\udea{^{;\al}}
\def\ddeb{_{;\b}}  \def\udeb{^{;\b}}

\def\naba{\nabla_{\a}}
\def\nabb{\nabla_{\b}}
\def\nabm{\nabla_{\m}}
\def\nabn{\nabla_{\n}}
\def\pmu{\partial_{\m}}
\def\pn{\partial_{\n}}
\def\p{\partial}

\def\bib#1{$^{\ref{#1}}$}

\def\al{\alpha} 
\def\b{\beta}
\def\g{\gamma}
\def\d{\delta}
\def\eps{\varepsilon}
\def\z{\zeta}
\def\h{\eta}
\def\th{\theta}
\def\k{\kappa}
\def\l{\lambda}
\def\m{\mu} 
\def\n{\nu} 
\def\x{\xi}
\def\r{\rho}
\def\s{\sigma}
\def\t{\tau}
\def\ph{\phi}
\def\ch{\chi}
\def\ps{\psi}
\def\om{\omega}
\def\G{\Gamma}
\def\D{\Delta}
\def\Th{\Theta}
\def\L{\Lambda}
\def\S{\Sigma}
\def\Ph{\Phi}
\def\Ps{\Psi}
\def\O{\Omega}

\def\ie{{\it i.e.}}
\def\cf{{\cal F}}
\def\ca{{\cal A}}
\def\cc{{\cal C}}
\def\cg{{\cal G}}
\def\cd{{\cal D}}
\def\cv{{\cal V}}
\def\cm{{\cal M}}
\def\co{{\cal O}}
\def\da{\dot{A}}
\def\db{\dot{B}}
\def\1{\dot{1}}
\def\2{\dot{2}}
\def\dd{\raisebox{+.3ex}{$\stackrel{\scriptstyle \leftrightarrow}{\partial}$}}
\def\etal{{\it et al.}\ }
\def\ie{{\it i.e. }}
\def\eg{{\it e.g. }}

\def\a{\`a }\def\o{\`o }\def\ii{\`\i{} }
\def\u{\`u  }\def\e{\`e }\def\ke{ch\'e }

\font\mybb=msbm10 at 12pt
\def\bb#1{\hbox{\mybb#1}}
\def\Z {\bb{Z}}
\def\R {\bb{R}}
\def\C {\bb{C}}
\def\H {\bb{H}}

\def\real{{\bb{R}}}
\def\rreal{{\bbb{R}}}
\def\rational{\bb{Q}}
\def\R4{\real^4}
\def\Ker{\mbox{Ker}}

\section{Introduction}
Many progresses have been made in recent years in the understanding of 
non--perturbative phenomena in globally supersymmetric (SUSY) gauge theories.
In a seminal paper \cite{sw}, Seiberg and Witten  calculated all the 
non--perturbative contributions to the holomorphic 
effective action for an $N=2$ Super Yang--Mills (SYM) theory.
The low--energy action can  in fact be expressed in terms of a 
unique holomorphic function $\cal F$, called the effective prepotential. 
Arguing that  the different phases of the theory
are described by a {\it polymorphic} prepotential,
Seiberg and Witten were able to translate
a set of physical conditions in terms of the number of singularities and the
monodromies of $\cal F$. From this knowledge, it was thus possible to 
reconstruct the entire prepotential. The only physical information concerning
non--perturbative effects fed to $\cf$ was the surviving quantum symmetry
of the theory which, for gauge group $SU(2)$, is $\Z_2$.
To check the reliability of this framework, the non--perturbative 
contributions to $\cf$ were computed directly  
in the Coulomb phase of the  theory 
(for gauge instantons of winding number one and two),
by using a saddle point approximation for certain correlators of the relevant
fields \cite{fp,dkm,ft}. The results were in agreement with those of 
Seiberg and Witten.
These successful checks raise a number of questions which are the motivations
of our investigation. The most compelling of them is:  How come that
a saddle point approximation, in which only quadratic terms are retained
in the expansion of the action, is able to give the correct result? Why
are higher--order corrections exactly zero? In the presence of SUSY a certain
number of welcome simplifications allow the 
computation of some 
Green's functions. The most remarkable simplification
is that these correlators are 
given by a constant
times the appropriate power of the 
renormalization group invariant (RGI) 
scale. This fact was exploited by Witten \cite{witten}
to argue that these computations are relevant for the determination
of a class of invariants of four dimensional manifolds. In the same paper
it was shown that  
in the topological twisted  version of $N=2$ SYM
the semiclassical expansion is exact.
Here we suggest that these guidelines could help us give an answer to the
problems previously raised.

\section{A Review of the Seiberg--Witten Model}
The Lagrangian density for the microscopic $N=2$ SYM theory, 
in the $N=2$ supersymmetric formalism is given by
\beq
\label{lagn=2n=2}
L= {1\over 16 \pi} \Im  \int \!  d^2 \theta d^{2} \tilde{\theta} \ 
\cf (\Psi ) \ \ .
\eeq
The chiral superfield $\Psi$, which describes the 
vector multiplet of the $N=2$ SUSY, transforms in the adjoint representation 
of the gauge group $G$ (which will be  $SU(2)$ from now on). 
Re--expressing the Lagrangian density in the $N=1$ formalism, 
we have 
\beq
\label{lagn=2}
L = {1\over 16 \pi}
\Im  \left[ \int \! d^2 \theta d^{2} \bar{\theta} 
K (\Phi , \bar\Phi , V ) + 
\int\! d^2 \theta f_{ab} (\Phi ) W^{a} W^{b} 
\right] \ \ ,
\eeq
where $a$, $b$ are indices of the adjoint representation of $G$.
The K\"{a}hler potential $K (\Phi , \bar\Phi , V ) $
and the holomorphic 
function  $f_{ab}(\Phi)$ 
are given, in terms of  $\cf$, by  
$K (\Phi , \bar\Phi , V )  =   
(\bar\Phi e^{-2V})^{a} {\p\cf/\p \Phi^a}$ and 
$f_{ab} (\Phi) =  {\p^2 \cf /\p \Phi^a \p\Phi^b}$.
The classical action 
is obtained by choosing for the holomorphic prepotential $\cf$ 
the functional form 
\beq
\label{preclass1}
\cf_{\rm cl} ( \Psi ) = { \tau_{\rm cl} \over 2} (\Psi^a \Psi^a  )\ \ ,
\eeq
where we define 
$
\tau_{\rm cl}= {\theta \over 2\pi} + {4\pi i \over g^2} 
$.
The classical action of the theory contains the scalar potential 
$
S_{\rm pot} =  \int d^{4} x\  \Tr  [ \phi , \phi^\dagger ]^2
$.
The most general (supersymmetric) classical vacuum configuration  is  then
\beq
\label{vaspf}
\ph_{0} = a \left( \Omega {\sigma_3 \over 2}\Omega^{\dagger} 
\right) \ \ , 
\ a \in \C \ \ , \ \ \Omega \in SU(2) \ \ .
\eeq
When $a \neq 0$ the $SU(2)$ gauge symmetry is spontaneously broken 
to $U(1)$.
The classical vacuum ``degeneracy''  
is lifted neither by perturbative nor by non--perturbative
quantum corrections \cite{Seibergnonren} so that one has a fully quantum
moduli space. 
The effective Lagrangian for the massless $U(1)$ fields  
$\Phi$,  $W_{\alpha}$,
will be the $U(1)$ version of Eq.~(\ref{lagn=2n=2}) and reads, 
in $N=1$ notation, 
\beq
\label{lagmassless}
L_{\rm eff} = 
{1\over 16 \pi}
\Im  \left[ \int\! d^2 \theta \cf^{\prime \prime} (\Phi ) W^{\alpha} 
W_{\alpha} + 
\int \! d^2 \theta d^{2} \bar{\theta} 
\bar\Phi \cf^{\prime} (\Phi )   
\right] \ \ .
\eeq
The low--energy dynamics are then governed by 
the effective prepotential $\cf (\Phi )$,
whose crucial property  is holomorphicity.
The effective coupling constant is given by
$
\tau (\Phi ) = \cf^{\prime \prime} (\Phi )
$.
The description of the low--energy dynamics in terms of 
$\Phi$ and $W_{\alpha}$ is 
not appropriate for all vacuum configurations.
The quantum moduli space  $\cm_{SU(2)}$
is better described in terms of the variable $a$ and its 
dual $a_{D}=\p_{a} {\cal F}$. When the gauge group is 
$SU(2)$, we can describe $\cm_{SU(2)}$
in terms of the 
gauge--invariant coordinate $u = <\Tr \phi^2 >$.   Then $\cm_{SU(2)}$
is the Riemann sphere with punctures at $u=\infty$
and $u=\pm \Lambda^2$, where $\L$ 
is the RGI scale 
in the normalization of Seiberg and Witten \cite{sw}.
At the classical level
\beq
\cf_{\rm cl} ( a )  = { \tau_{\rm cl} \over 2} a^2 \ \ ;
\eeq
however perturbative as well as non--perturbative effects modify the 
expression of the prepotential, so that 
\beq
\cf (a) = \cf_{\rm pert} (a)  +\cf_{\rm np} (a) 
\ \ .
\eeq
The perturbative term has been calculated by Seiberg  
\cite{sei}
and is  
\beq
\cf_{\rm pert}  (\Psi ) =  {i\over 2\pi}\Psi^2  \ln 
{\Psi^2 \over M^2} \ \ ,
\eeq
where $M$ can be fixed   
by assigning the value of  the coupling constant at some subtraction 
point. 
The $R$--symmetry of the theory constrains the  non--perturbative 
prepotential to be 
\beq
\label{nat.01}
\cf_{\rm np} (a) = 
\sum_{k=1}^{\infty} \cf_k \left( {\Lambda \over a }\right)^{4 k}  a^2 
\ \ ,
\eeq
and similarly
\beq
\label{espditrfiquad}
u(a) ={1\over 2} a^2 +  \sum_{k=1}^{\infty} \cg_k 
\left({\Lambda\over a}\right)^{4k } a^2 
\ \ . 
\eeq
The 1--instanton contribution to $u(a)$  was found to be 
\cite{fp,ft}
\beq
\label{cg1pv}
<\Tr \phi^2 >_{k=1} = {\Lambda_{\rm PV}^4 \over  g^4 a^2}
\ \ 
\eeq
where 
$\Lambda_{\rm PV}=\L/\sqrt{2}$ is the Pauli--Villars RGI scale, 
which naturally arises when performing instanton calculations.
Matone's relation \cite{marco}
expresses then the $\cf_k$'s as functions of the $\cg_k$'s, 
\beq 
\label{ca.13}
2 i \pi k \cf_k = \cg_k 
\ \ .
\eeq
By making some hypotheses on the structure of
the moduli space and on the monodromies of $\tau$ around its  singularities, 
Seiberg and Witten obtained the expressions of $a(u)$ and 
$a_{D}(u)$, 
\beqa
\label{adiu}
a(u)&=&\frac{\sqrt{2}}{\pi}\int_{-\L^{2}}^{\L^{2}} d x\ \frac{\sqrt{x-u}}
{\sqrt{x^2-\L^{4}}}
\ \ ,
\\
\label{addiu}
a_{D}(u)&=& \frac{\sqrt{2}}{\pi}
\int_{\L^{2}}^{u} dx \ \frac{\sqrt{x-u}}
{\sqrt{x^2-\L^{4}}}
\ \ .
\eeqa
From Eqs.~(\ref{adiu}), (\ref{addiu}) one can derive the ${\cal F}_{k}$'s 
and check whether they agree with those computed via instanton 
calculations, which we study in the next Section. 
\section{Instanton Calculus in SUSY Gauge Theories}
We now briefly review the strategy 
to perform 
semiclassical computations in supersymmetric gauge theories,
in the context of the constrained instanton method \cite{aff}.

We expand the action functional 
around  a properly chosen field configuration, which is the 
solution of the  equations 
\beq
\label{pero}
D_{\mu} (A) F_{\mu \nu} = 0 \ \ ,
\eeq
\beq
\label{classphi}
D^{2} (A) \phi_{\rm cl} = 0 \ \ , \ \ \lim_{|x| 
\rightarrow \infty}
\phi_{\rm cl} \equiv \phi_\infty = a \frac{\sigma_{3}}{2i}\ \ . 
\eeq
The boundary condition in Eq.~(\ref{classphi}) is dictated by 
Eq.~(\ref{vaspf}).
Equation  (\ref{pero}) admits instanton solutions.
When $a=0$, Eq.~(\ref{classphi}) has only the 
trivial solution 
$\phi =0$. On the other hand, when $a \neq 0$, we shall decompose
the fields $\phi$, $\phi^{\dagger}$ as
\beq
\begin{array}{c c}
\phi= \phi_{\rm cl} + \phi_{Q} \ \ , & 
\ \ \phi^{\dagger}= (\phi_{\rm cl})^{\dagger}
+ \phi_{Q}^{\dagger} \ \ ,
\end{array}
\eeq
and integrate over the quantum fluctuations $\phi_{Q}$, 
$\phi_{Q}^{\dagger}$.
We want to remark that Eqs.~(\ref{pero}), (\ref{classphi}) 
are just approximate  saddle point 
equations of the $N=2$ action functional.  
This is a tricky point of the constrained instanton method. 

The integration over the bosonic zero--modes can be traded for
an integration over the collective coordinates, at the cost of 
introducing  the corresponding Jacobian.
The existence of fermion zero--modes  is the 
way by which Ward identities  related to the 
group of   chiral symmetries of the theory
come into play.  
When $a=0$,  the  anomalous  $U(1)_{R}$ symmetry
\beq
\begin{array}{c c}
\lambda \longrightarrow e^{i \alpha} \lambda \ \ ,
& \ \ 
\phi \longrightarrow e^{2 i \alpha}  \phi
\end{array}
\eeq
and gauge invariance  allow 
a nonzero result    for the 
Green's functions   with $n$ insertions of
the gauge--invariant quantity  $(\phi^{a } \phi^{a })(x)$ only 
when $n = 2 k $.
These correlators possess the right operator insertions 
needed   to saturate 
the integration over the  Grassmann parameters and, 
by supersymmetry,  they  are also 
position--independent.  
On the other hand, when  $a\neq 0$,   
the correlator $\left\langle \phi^{a} \phi^{a} \right\rangle$
admits a complete expansion in terms of instanton contributions.

Fermion zero--modes are found by solving the equation 
$
D_{\mu} \bar{\sigma}^{\dot{\alpha}\beta}_{\mu} 
\lambda_{\beta \dot{A}} = 0 
$, where $\dot{A}=1,2$ is the supersymmetry index.
For instantons of winding number $k=1$
 the whole set of solutions  
of this equation is obtained via SUSY and superconformal 
transformations, which yield 
\beq
\lambda_{\alpha \dot{A}}^{a} = 
\frac{1}{2}F_{\mu \nu}^{a}
(\sigma_{\mu \nu})_{\alpha}^{\ \beta}
 \zeta_{\beta \dot{A}} \ \ ,
\label{g.5}
\eeq
where 
$\zeta = \xi + (x-x_{0})_{\mu} \sigma_{\mu} 
\bar{\eta}/\sqrt{2} \rho $, 
$\xi$, $\bar{\eta}$ being two arbitrary quaternions
of Grassmann numbers.  

The correct fermionic integration measure is given 
by the inverse of the determinant of the matrix whose 
entries are the scalar products of the fermionic zero--mode
eigenfunctions and it reads 
$d^{4} \xi d^{4} \bar{\eta} \left( \frac{g^2}{32 \pi^2} \right)^{4}$, 
where $d^{4} \xi d^{4} \bar{\eta} \equiv
d^{2} \xi_{\dot{1}} d^{2} \xi_{\dot{2}}
d^{2} \bar{\eta}_{\dot{1}}  d^{2} \bar{\eta}_{\dot{2}}  $.

As an example of instanton calculation we now consider 
the correlator $<\phi^{a} \phi^{a} >$ in the semiclassical approximation 
around an instanton background of winding number $k=1$,  
\beqa \label{k=1}
&&<\phi^{a} \phi^{a} > = 
\nonumber \\
&&\int d^{4}   x_{0} d \rho \ 
\left(  \frac{2^{10} \pi^{6} \rho^{3}}{g^{8}} \right)
e^{-\frac{8 \pi^2}{g^2} - 4 \pi^{2} |a|^{2} \rho^{2}}
\nonumber \\
&&    
\int [ \delta Q  \delta \lambda ] \delta \bar{\lambda}  
\delta \phi^{\dagger}_{Q} \delta \phi_{Q}
\delta \bar c  \delta c  
\nonumber \\
&&
\exp \biggl[ - S_H[\phi_Q,\phi_Q^\dagger,A^{cl}]-S_F[\lambda,
\bar\lambda,A^{cl}]+
\nonumber \\
&& - \frac{1}{2} \int d^4x \ Q_{\mu} M_{\mu \nu} Q_{\nu}
- \int d^4x \ \bar c D^2 (A^{cl}) c \biggr]
\nonumber \\
&&\int
d^{4} \xi d^{4} \bar{\eta} 
\left( \frac{g^2}{32 \pi^2} \right)^{4}
\nonumber\\
&&\exp \left[ -S_{\rm Y} [ \phi_{\rm cl} + \phi_{Q}, 
 (\phi_{\rm cl})^{\dagger} + \phi_{Q}^{\dagger}  , 
\lambda^{(0)},
\bar\lambda=0 ] 
\right] 
\nonumber \\ 
&&(\phi_{\rm cl} + \phi_{Q})^{a} ( \phi_{\rm cl} + 
\phi_{Q})^{a}(x) \ \ \ .
\eeqa
Let us now explain where 
the different terms in Eq.~(\ref{k=1})
come from: 
\begin{itemize}
\item[{\bf 1.}]
$ d^{4}   x_{0} d \rho \ 
\left(\frac{2^{10} \pi^{6} \rho^{3}}{g^{8}} \right)$
is the bosonic measure after the integration over
$SU(2) / \Z_{2}$ global rotations in color space has been
performed.  $x_{0}$ and 
$\rho$ are the center and the size of the instanton.
\item[{\bf 2.}]
$S_H[\phi_{cl},(\phi_{cl})^\dagger,A^{cl}]=4 \pi^{2} |a|^{2} 
\rho^{2}$,
is the contribution of the classical Higgs action.
\item[{\bf 3.}]
The third and fourth lines  include the quadratic
approximation of the different 
kinetic operators for the quantum fluctuations 
of the fields, and the 
symbol $[\delta \lambda \delta Q]$ denotes integration 
over nonzero--modes.  $\bar c$ and $c$ are the usual ghost fields, 
$\int d^4x \ \bar c D^2 (A^{cl}) c$ 
being the corresponding term in the action.

\item[{\bf 4.}]
$S_{\rm Y} \left[ \phi ,  \phi^{\dagger}  , \lambda^{(0)},
\bar\lambda=0 \right] $
is the Yukawa action calculated with the complete expansion 
of the fermionic  fields replaced by their   
projection over the zero--mode 
subspace. 
It reduces to 
$\sqrt{2} g  \epsilon^{abc} 
\int \phi^{a \dagger} (\lambda_{\dot{1}}^{(0) b} 
\lambda_{\dot{2}}^{(0) c})$.
\end{itemize}
After the  integration over $\phi$, $\phi^{\dagger}$ and
the nonzero--modes,  the $\phi_{Q}$   insertions
get replaced by $\phi_{\rm inh}$, where
\beq 
\phi_{\rm inh}^{a}= \sqrt{2} g  \epsilon^{bdc} 
[( D^{2})^{-1}]^{ab} (\lambda_{\dot{1}}^{(0) d}  
\lambda_{\dot{2}}^{(0) c}) \ \ ,
\label{fiino}
\eeq
and the determinants of the various kinetic operators cancel
against each  other.
The r.h.s. of Eq.~(\ref{k=1}) now reads 
\beqa
&&\Lambda_{\rm PV}^{4}\int d^{4}   x_{0} d \rho \ 
\left(\frac{2^{10} \pi^{6} \rho^{3}}{g^{8}} \right)
e^{- 4 \pi^{2} |a|^{2} \rho^{2}}    
\nonumber \\
&&\int d^{4} \xi d^{4} \bar{\eta}
 \left( \frac{g^2}{32 \pi^2} \right)^{4}
 \
\exp \left[ - \sqrt{2} g  \epsilon^{abc} 
\int \phi_{cl}^{\dagger} (\lambda_{\dot{1}}^{(0) b} 
\lambda_{\dot{2}}^{(0) c}) \right] 
\nonumber \\ 
&&(\phi_{\rm cl} + \phi_{\rm inh })^{a} 
 ( \phi_{\rm cl} + \phi_{\rm inh})^{a}(x)
 \ \ \ ,
\label{correlatore}
\eeqa
where 
$\Lambda_{\rm PV}^{4}=\mu^{8-\frac{1}{2}(4+4)} e^{- \frac{8 \pi^2}{g^2}}$. 
$\mu$ comes from the Pauli--Villars
regularization of the determinants and the exponent is
$b_1k=(n_B-n_F/2)$ where $n_B,n_F,b_1$ are the number of bosonic,
fermionic zero--modes and the first coefficient of the 
$\beta$--function of the theory. 

The Yukawa action does {\em not} contain  
the Grassmann parameters of the zero--modes coming
from   SUSY transformations. 
As a consequence 
the only 
nonzero contributions are  obtained by  picking out
the terms in the $\phi_{\rm inh}$ insertions 
which contain the SUSY solutions
of the Dirac equation.  
This amounts to say%
\footnote{This property generally holds for all the 
multi--instanton contributions to $u(a)$.}
\beq
(\phi_{\rm cl} + \phi_{\rm inh })^{a}  
( \phi_{\rm cl} + \phi_{\rm inh})^{a}
\longrightarrow
 - \xi_{\dot{1}}^{2} \xi_{\dot{2}}^{2} 
(F_{\mu \nu}^{a} F_{\mu \nu}^{a})  .
\label{fireplace}
\eeq
Equation (\ref{k=1}) now becomes
\beqa
<\phi^{a} \phi^{a} > \ \ = 
&& \Lambda_{\rm PV}^{4}\int d \rho \ 
\left(\frac{2^{10} \pi^{6} \rho^{3}}{g^{8}} \right)
e^{- 4 \pi^{2} |a|^{2} \rho^{2}}    
\nonumber \\
&& \int d^{4} x_{0}\ (F_{\mu \nu}^{a} F_{\mu \nu}^{a})  
\nonumber\\
&&\frac{g^2}{2} (a^{\ast})^{2}
\int d^{4} \xi   \left( \frac{g^2}{32 \pi^2} \right)^{2} \ 
\xi_{\dot{1}}^{2} \xi_{\dot{2}}^{2} 
 \ \ \ .
\label{g1}
\eeqa
We can then immediately integrate over
$x_{0}$ remembering that 
$\int d^{4} x \  F_{\mu \nu}^{a} F_{\mu \nu}^{a} = 
32 \pi^2 / g^2$. 
The remaining integrations over $\xi$ and
$\rho$ in Eq.~(\ref{g1}) are trivial and yield
\cite{fp,ft}
\beq
<\phi^{a} \phi^{a} > = \frac{2}{g^4}\frac{\Lambda_{\rm PV}^{4} }{a^2} \ \ .
\eeq
This result agrees with the Seiberg--Witten prediction.

\section{Topological Yang--Mills Theory and Instanton Moduli Spaces}
The relationship between supersymmetric and topological theories
shows up when one observes that in the former there exists 
a class of position--independent Green's functions. 
When formulated 
on a generic manifold $M$ by redefining 
the generators of the Lorentz group in a suitably twisted fashion,
the $N=2$ SYM theory
gives rise to the so--called Topological Yang--Mills
theory (TYM) \cite{witten}. {\it All} the observables, included 
the partition function itself,  are in this case 
topological invariants, in the sense that they are independent of 
the metric on $M$.

With respect to the twisted Lorentz group, SUSY charges decompose
as a scalar $Q$, an antisymmetric tensor $Q_{\mu\nu}$ and a vector
$Q_\mu$:
\beqa
&&\bar{Q}_{\dot\alpha}^{\dot A}\rightarrow Q \oplus Q_{\mu\nu} \ \ , \nonumber\\
&&Q_{\alpha}^{\dot A}\rightarrow Q_\mu \ \ .
\label{tw-ch}
\eeqa
Moreover, the twist transforms the $N=2$ SYM fields as 
\beqa
&&A_\mu \rightarrow A_\mu \ \ , \nonumber\\
&&\bar{\lambda}_{\dot\alpha}^{\dot A}\rightarrow \eta \oplus \chi_{\mu\nu}
\ \ , \nonumber\\
&&\lambda_{\alpha}^{\dot A} \rightarrow \psi_\mu \ \ , \nonumber\\
&&\phi \rightarrow \phi \ \ ,
\label{tw-fi}
\eeqa 
where the anticommuting fields $\eta,\chi_{\m\n},\psi_{\m}$ are 
respectively a scalar, a self--dual two--form and a vector.
The scalar supersymmetry charge of TYM plays a major r\^{o}le, in that it is 
preserved on any (differentiable) four--manifold
$M$, and has the crucial property of being nilpotent modulo gauge 
transformations.
This allows one to interpret it as a BRST--like charge. 
Actually, in order to have a strictly
nilpotent BRST charge, one needs to include gauge transformations
with the appropriate ghost $c$ \cite{bs}.
The BRST transformations are then
\beqa\label{BRST}
&&sA=\psi-Dc\ \ ,\nonumber\\
&&s\psi=-[c,\psi]-D\phi\ \ ,\nonumber\\
&&sc=-[c,c]+\phi\ \ ,\nonumber\\
&&s\phi=-[c,\phi]\ \ .
\eeqa
This algebra can be read as the definition and the Bianchi identities
for the curvature $\hat{F}=F+\psi+\phi$ of the connection $\hat{A}=A+c$ of 
the universal
bundle $P\times {\cal A}/{\cal G}$ ($P, {\cal A}, {\cal G}$ are 
respectively the principal bundle over $M$, the space of connections and
the group of gauge transformations).
The exterior derivative on the base manifold $M\times {\cal A}/{\cal G}$
is given by $\hat{d}=d+s$. 

The TYM action can be interpreted as a 
pure gauge--fixing action which localizes the 
universal connection to 
\beq
\hat{A} = A+c = U^\dagger (d+s) U \ \ .
\label{apiuc-adhm}
\eeq
$A$ is then an anti--self--dual (ASD) connection which we have 
written in the ADHM formalism%
\footnote{We use the definitions and conventions of Sec. II of \cite{ft}.}.
Once $\hat{A}$ is given, the components $F, \psi, \phi$ of
$\hat{F}$ are in turn determined. $F$ is anti--self--dual ($F=F^-$), and 
$\psi$ is an element of the 
tangent bundle $T_A{\cal M^-}$, where ${\cal M}^-$ is 
the instanton moduli space. Moreover, the scalar field $\phi$ 
is the solution to the equation 
$D^2 \phi = [\psi ,  \psi]$. The explicit expression  for $\phi$
is then given by the twisted
version of Eq.~(\ref{fiino}). 
For the following discussion it is important to point out 
that the $\phi$ field has
trivial boundary conditions ($\phi=0$) at spatial infinity.

In this geometrical framework, the BRST operator $s$ has a
very nice explicit realization as the exterior derivative on ${\cal M}^-$.
This leads us to compute correlators of $s$--exact operators as integrals 
of forms on $\partial {\cal M}^-$ \cite{bftt}.
For example,  we can write 
\beq
\Tr \phi^2 = s K_{c}\ \ , \ \ 
K_c = \Tr \left(csc + {2\over 3} ccc \right) \ \ ,
\eeq
an expression which parallels the well--known relation
\beq
\Tr F^2 = s K_{A}\ \ , \ \ 
K_A =
\Tr \left(AdA + {2\over 3} AAA\right) \ \ .
\eeq
For winding number $k=1$,  the top form on the 
(eight--dimensional) instanton moduli space is 
$\Tr \phi^2 (x_1)  \Tr \phi^2  (x_2)$, and one can compute  
\cite{bftt}
\beq
\int_{{\cal M}^-} \Tr \phi^2 \Tr \phi^2 = 
\int_{\partial {\cal M}^-} \Tr \phi^2 K_c = {1\over 2}
\ \ .
\eeq

\section{Topological Aspects of the Seiberg--Witten Model}
From the comparison between the Seiberg--Witten ansatz for the 
$N=2$ low--energy effective action and  explicit instanton calculations,
we learn two important lessons.
On one hand, this comparison provides us with 
a consistency check of the scenario proposed by Seiberg and Witten
\cite{sw}. On the other hand, 
it strongly suggests that the semiclassical approximation
around the instanton background saturates the non--perturbative sector.
This calls for an explanation.
A key property of the TYM theory is the exactness of the semiclassical 
limit \cite{witten}: 
we are then naturally  led  to explore the Coulomb phase of 
the $N=2$ SYM theory starting from its topological twisted counterpart.

The first problem one must face is that nontrivial boundary conditions
for the scalar field are not compatible with the BRST algebra in 
Eq.~(\ref{BRST})%
\footnote{Recall that $c$ localizes to $U^{\dagger}s U$.}. 
This is because a nonzero v.e.v. for $\phi$
implies the existence of a (nonzero) central charge $Z$ in the SUSY algebra 
which acts on the fields as a $U(1)$ transformation
with gauge parameter $\phi$.
This new symmetry 
has to be included in an appropriate extension of the BRST operator, 
and it is implemented through the introduction of a new 
{\it global} ghost field $\L$. 
The resulting algebra is then \cite{bftt}
\beqa\label{EBRST}
&&sA=\psi-D(c+\L)\ \ ,\nonumber\\
&&s\psi=-[c+\L,\psi]-D\phi\ \ ,\nonumber\\
&&s(c+\L)=-[c+\L,c+\L]+\phi\ \ ,\nonumber\\
&&s\phi=-[c+\L,\phi]\ \ .
\eeqa
Once the universal connection is projected onto 
Eq.~(\ref{apiuc-adhm}) by the gauge--fixing TYM action,
the above extended algebra  includes  scalar fields 
with nonvanishing boundary conditions. Therefore, 
we can see that in this picture  
the field configurations dictated by the constrained instanton
method (see Eqs.~(\ref{pero}), (\ref{classphi})) 
naturally come into play, without 
resorting to any approximation procedure.  

The TYM action gets now a nonzero boundary contribution from the term
\beq
\label{DKM}
S_{\rm inst}= \int_{R^4} d^4x \ 
\partial^{\mu} s \Tr ( \phi^{\dagger} \psi_{\mu} )
\ \ .
\eeq
The explicit ADHM expressions for the fields in Eq.~(\ref{DKM})
can be derived from the  extended algebra  in Eq.~(\ref{EBRST})
starting from the universal connection $U^\dagger (d+ s) U$%
\footnote{In this context, the global ghost $\L$ is necessary to ensure 
invariance under local reparametrizations in the moduli space.}.
It is easy to see \cite{bftt} that,
when inserted into Eq.~(\ref{DKM}),  
they  yield the multi--instanton action \cite{dkm} 
for the Seiberg--Witten model.
This provides a natural and simplifying  framework for further studies 
of non--perturbative effects in $N=2$ theories.

\section*{Acknowledgements}
It is a pleasure to thank D. Anselmi, C.M. Becchi, S. Giusto,
C. Imbimbo,  M. Matone, 
G.C. Rossi and S.P. Sorella for many stimulating discussions.


\section*{References}
\begin{thebibliography}{99}
\bibitem{sw}N. Seiberg and E. Witten, 
\Journal{\NPB}{431}{19}{1994}.

\bibitem{fp}D. Finnell and P. Pouliot,
\Journal{\NPB}{453}{225}{1995}. 


\bibitem{dkm}N. Dorey, V.V. Khoze and M.P. Mattis, 
\Journal{\PRD}{54}{2921}{1997}.

\bibitem{ft}F. Fucito and G. Travaglini, \Journal{\PRD}{55}{1099}{1997}.

\bibitem{witten} E. Witten, 
{\em Commun.~Math.~Phys.}  {\bf 117}, 353 (1988).


\bibitem{Seibergnonren} N. Seiberg, 
\Journal{\PLB}{328}{469}{1993}. 

\bibitem{sei}  N. Seiberg, 
\Journal{\PLB}{206}{75}{1988}. 

\bibitem{marco}  M. Matone, 
\Journal{\PLB}{357}{342}{1995}. 

\bibitem{aff}
I. Affleck, \Journal{\NPB}{191}{429}{1981}; 
I. Affleck, M. Dine and N. Seiberg, 
\Journal{\PRL}{51}{1026}{1983}; 
V. Novikov, M. Shifman, A. Vainshtein and
V. Zakharov,  \Journal{\NPB}{260}{157}{1985}. 


\bibitem{bs}L. Baulieu and I.M. Singer, 
{\em Nucl. Phys. Proc. Suppl.} {\bf B5}
12 (1988). 


\bibitem{bftt}  D. Bellisai, F. Fucito, A. Tanzini and 
G. Travaglini, {\it in preparation}. 

\end{thebibliography}
\end{document}